# High-Density Monocular 3D Particle Image Velocimetry by Wavefront Shaping and Deep Learning

Dafei Xiao[1], Jibu Tom Jose[3], Amit Parizat[1], Reut Orange Kedem[4], Josué Sznitman[1], Omri Ram[3], Yoav Shechtman[1,2,4*]

[1] Faculty of Biomedical Engineering, Technion - Israel Institute of Technology, Haifa, Israel
[2] Faculty of Electrical and Computer Engineering, Technion - Israel Institute of Technology, Haifa, Israel
[3] Faculty of Mechanical Engineering, Technion - Israel Institute of Technology, Haifa, Israel
[4] Russell Berrie Nanotechnology Institute, Technion–Israel Institute of Technology, Haifa, Israel
*yoavsh@technion.ac.il

## Abstract

Three-dimensional (3D) Particle Image Velocimetry (PIV) measures flow velocities by imaging laser-illuminated tracer particles seeded in a fluid, and is widely used in both academia and industry. Many applications require a compact setup for optical accessibility, ideally with a single camera, while also demanding high seeding densities for accurate velocimetry. These requirements, however, are typically incompatible: monocular methods break down at high densities because of particle-image overlap; high-density measurements instead generally rely on multi-camera tomographic systems. Here, we introduce Point-spread-function-Engineering Training-based PIV (PET-PIV), a compact monocular 3D velocimetry approach that resolves this long-standing compactness-density trade-off through a minimal optical modification and deep-learning-based algorithms. PET-PIV requires only the insertion of a single thin phase mask to an otherwise conventional PIV setup, beneath the objective in microscopy or at the lens iris in macro-scale imaging, and calibrates the resulting imaging system in situ, making the approach straightforward to implement and readily scalable. Computationally, PET-PIV operates in two complementary regimes: a tracking/Lagrangian mode that localizes and links individual particles, and, more importantly, a field/Eulerian mode for ultra-high densities which directly reconstructs 3D velocity fields from 2D image sequences. In realistic experimental validations using a state-of-the-art tomographic PIV (TPIV) system, PET-PIV demonstrates strong agreement with ground-truth references, with correlation coefficients (CC) exceeding 0.97, alongside an order-of-magnitude improvement in computational speed. Notably, PET-PIV enables monocular 3D macro-scale PIV at densities previously achievable only with multi-camera setups, dramatically expanding the applicability of 3D velocimetry in space-constrained and optically restricted environments.

## Introduction

The measurement of velocity fields in fluid flows is essential for experimental validation of theoretical fluid mechanics, and for the design and optimization of fluidic systems, spanning scales from microfluidic devices[1] to aircrafts[2–4]. Among existing techniques, 3D particle image velocimetry (PIV), which uses laser-illuminated passive tracer particles seeded in the flow, has become the dominant approach. One of the most widely applied and commercialized 3D PIV techniques is tomographic PIV (TPIV)[5], which uses multiple views and cameras (typically four to six) for stereo-vision-based 3D imaging, providing 3-dimensional 3-component (3D-3C) flow reconstruction. Despite its success, TPIV's multi-camera configuration imposes practical limitations. Most notably, system bulkiness can severely restrict optical access[6]. In addition, it requires substantial memory storage and is computationally slow. Hardware costs are significant, especially when high-speed cameras are needed, and calibration and alignment procedures are both expertise-intensive and time-consuming. These constraints have motivated persistent efforts to develop compact monocular 3D PIV techniques.

Several single-camera 3D imaging strategies have been explored for PIV. For example, light field PIV employs a microlens array, placed either at the native image plane[7] or at a Fourier plane[8], to enable the joint sampling of spatial and angular information of light rays. Holographic PIV[9,10] records phase-encoded holograms, which can be processed to reconstruct the complex optical field and 3D particle distribution. In contrast, structured-illumination PIV relies on depth-encoded illumination patterns, such as rainbow-colored[11,12] and intensity-gradient[13] illumination, to generate particle images with depth-discriminative features.

Another particularly attractive direction relies on point spread function (PSF) engineering, also referred to as coded aperture or wavefront shaping. In these approaches, a mask inserted at the aperture of the optical system modifies the PSF, i.e., the particle image, so that its shape efficiently encodes depth information (the axial position of the particle). Representative examples include three pinhole apertures[14,15] which encode depth through the lateral positioning of three spots, and cylindrical lenses[16–18] which exploit PSF ellipticity. A decade ago, a Tetrapod phase mask[19] was used for sparse particle tracking velocimetry (PTV) in microfluidic flows[20].

Although these monocular 3D PIV methods improve compactness and optical accessibility, they are generally limited to sparse particle tracking, and struggle at high particle seeding densities[21], which are critical for accurate velocimetry. Furthermore, many require nontrivial optical modifications, such as 4f relay systems, special cameras, customized illumination setups, as well as complex calibration procedures, limiting their practical deployment. These limitations motivate the need for a monocular 3D PIV framework that preserves compactness while operating robustly across particle densities and imaging scales, thereby enabling broadly accessible volumetric flow measurements with the potential to transform experimental fluid mechanics across a wide range of scientific and engineering disciplines.

Here, we present PET-PIV, a general framework that resolves the long-standing trade-off between optical accessibility and seeding density, and operates robustly across densities and imaging scales. At the hardware level, the required optical modification is extremely minimal and cost-effective: in microscopy, a phase mask is inserted beneath the objective, whereas at the macroscale it is mounted at the aperture plane of the lens. Computationally, PET-PIV operates in two complementary modes. In tracking mode, it performs particle-resolved tracking velocimetry, where, to address the localization challenges caused by PSF overlap, we employ a deep neural network (LocNet) trained on realistic synthetic data generated from an in-situ calibrated PSF model. Experimental validation against a commercial TPIV system demonstrates quantitative agreement in both swirl and jet flows at a Reynolds number (Re) regime of 1000. At ultra-high seeding densities, PET-PIV switches to field mode, which exploits the global variations in image properties induced by the flow, rather than resolving individual particles, and uses a second neural network (VelNet) supervised by TPIV-aligned ground truth to directly reconstruct 3D-3C velocity fields from 2D image sequences. Experiments show that the monocular system operates at particle densities up to 0.4 particles per cubic millimeter (PPCM) and achieves over one-order-of-magnitude improvement in processing speed without loss of measurement accuracy, compared to multi-camera TPIV. The combination of compactness and high-density measurement capability makes PET-PIV a practical alternative to multi-camera volumetric velocimetry and, perhaps most importantly, uniquely positions it for velocimetry in optically constrained environments.

## PET-PIV principle

We introduce PET-PIV, a monocular volumetric particle imaging method, that combines a simple implementation of PSF engineering with learning-based reconstruction to overcome the limitations of previous approaches. Conventional tomographic reconstruction of a particle field, such as typically used for tomographic particle image velocimetry, employs a powerful light source to illuminate particle-seeded flow (Fig. 1), and multiple cameras (C1-C4 in Fig. 1) viewing the flow from different perspectives. Each camera operates with a stopped-down aperture to produce a large depth of field, ensuring that particles throughout the illuminated volume remain approximately in focus (Fig. 1, bottom left). In contrast, PET-PIV employs a single camera (C0 in Fig. 1) with a diffractive optical element, namely, a Tetrapod phase mask[19], mounted at an opened lens aperture. This coded aperture engineers the PSF, i.e., the particle image, such that its shape efficiently encodes axial position. As shown in the C0-linked image and the five PSFs at different axial positions in Fig. 1, the PSF footprint broadens and exhibits depth-dependent elongation.

When processing the image sequences, PET-PIV can use one of two complementary modes, depending on particle seeding density and measurement requirements. In tracking mode, follows a localization–linking–interpolation pipeline (Fig. 1). During localization, image frames are processed by a deep neural network (LocNet, adapted from DeepSTORM3D[22,23]), which outputs particle coordinates for each frame. LocNet is trained on realistic synthetic images generated from an *in-situ* PSF model. The model is calibrated using spare-particle data, with the known illumination thickness (axial span) as a physical constraint. During linking, a four-frame algorithm[24] is used to construct particle trajectories, followed by trajectory smoothing and velocity calculation. The resulting velocity measurements are mapped onto a Eulerian grid using a constrained cost minimization (CCM)-based interpolation method[25]. Tracking-based results are validated against ground-truth TPIV data processed using the STB[26].

At ultra-high seeding densities, particle-localization and subsequent linking becomes impractical, thus the image sequence is directly mapped to volumetric velocity fields using a second neural network, VelNet (Fig. 1). The network takes five consecutive frames and outputs three velocity components (U, V, W), notably, without performing particle localization and linking. During training, ground-truth supervision is obtained from TPIV data processed with a correlation-based algorithm, following precise geometric alignment among cameras C0–C5. Once trained, PET-PIV operates entirely with the single camera C0.

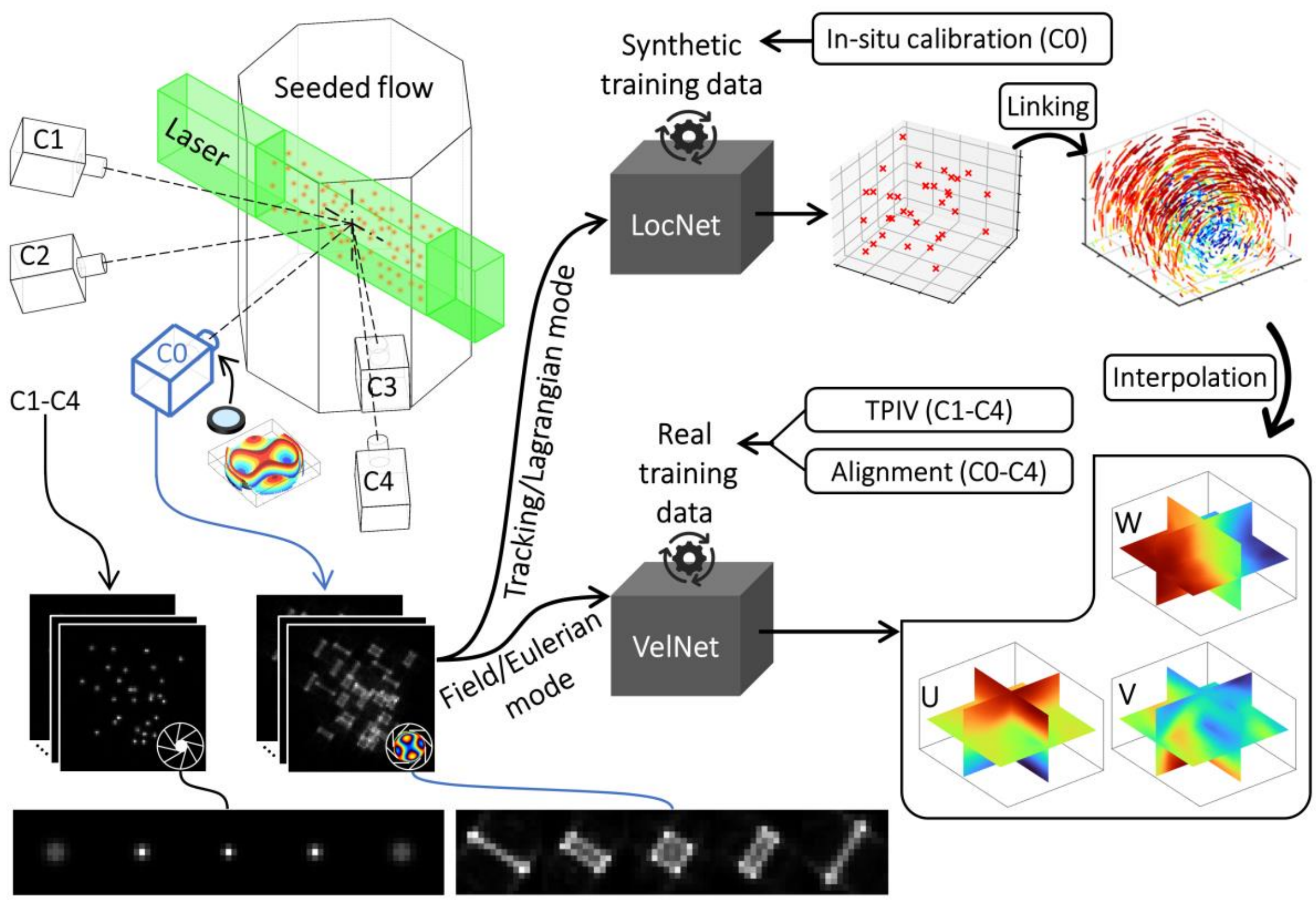

Fig. 1: Schematic of PET-PIV. A single camera (C0) equipped with a Tetrapod phase mask at the lens aperture generates depth discriminative PSFs (see the five C0-linked PSFs at different axial positions), whereas the TPIV cameras (C1-C4) operate with stopped-down apertures to produce identical in-focus PSFs across the illuminated volume of the seeded flow (see the five PSFs linked to C1-C4). In tracking mode, C0 image sequences are processed through localization–linking–interpolation pipeline: particle positions are obtained using LocNet, trained on synthetic images generated from an in-situ calibrated PSF model, followed by linking and velocity interpolation to reconstruct volumetric velocity fields. In field mode, VelNet directly predicts the 3D velocity field from the image sequence, with training supervised by TPIV-aligned ground truth.

## Tracking mode of PET-PIV in microscopy

To demonstrate the usability of PET-PIV across length and flow scales, we validate the tracking mode of PET-PIV in both micro and macro applications. Starting with a microfluidic configuration, the phase mask is compactly positioned beneath the objective via an accessory slot of the microscope (Fig. S1a). We first measure a steady flow (Supplementary Video S1) in a passive microfluidic chip (Fig. S1b) with a cross-sectional area change (Fig. 2a) where the flow is driven by capillary forces arising from specially designed microstructures[27]. The channel height is ~40 µm on the left and ~100 µm on the right, and the Re is on the order of 0.001. A representative frame is shown in Fig. 2b, with the calibrated PSF model displayed above. In region c, LocNet localizations are illustrated: x-y positions are shown by red marks on the measured image (Fig. 2c, top), while axial (z) positions are visualized through PSF shapes at the corresponding x-y locations (Fig. 2c, bottom simulated “sim” image). Localizations from the full video are then linked through a four-frame algorithm[24] to construct particle trajectories (Fig. 2d and Supplementary Video S2, colored by velocity magnitude), from which velocities are computed. Velocity profiles are extracted along the blue and orange lines in Fig. 2d. The distributions of the streamwise (x) velocity component U along depth (z) axis are shown in Fig. 2e, demonstrating the expected parabolic profiles.

Next, we image an unsteady flow (Supplementary Video S3) in a straight microchannel with a height of 20 µm (Fig. 2f, see calibrated PSF and a representative frame in Fig. S1e). The flow speed and direction are controlled by a pump (Fig. S1f), such that the fluid initially flows to the right and gradually reverses direction, with Re on the order of $10^{-3}$. The time-varying flow is characterized by the distribution of the streamwise (x) velocity component U along z axis (Fig. 2g). A representative 2D U-velocity profile in the y–z plane at t = 0.2 s is presented in Fig. 2h (full result in Supplementary Video S4).

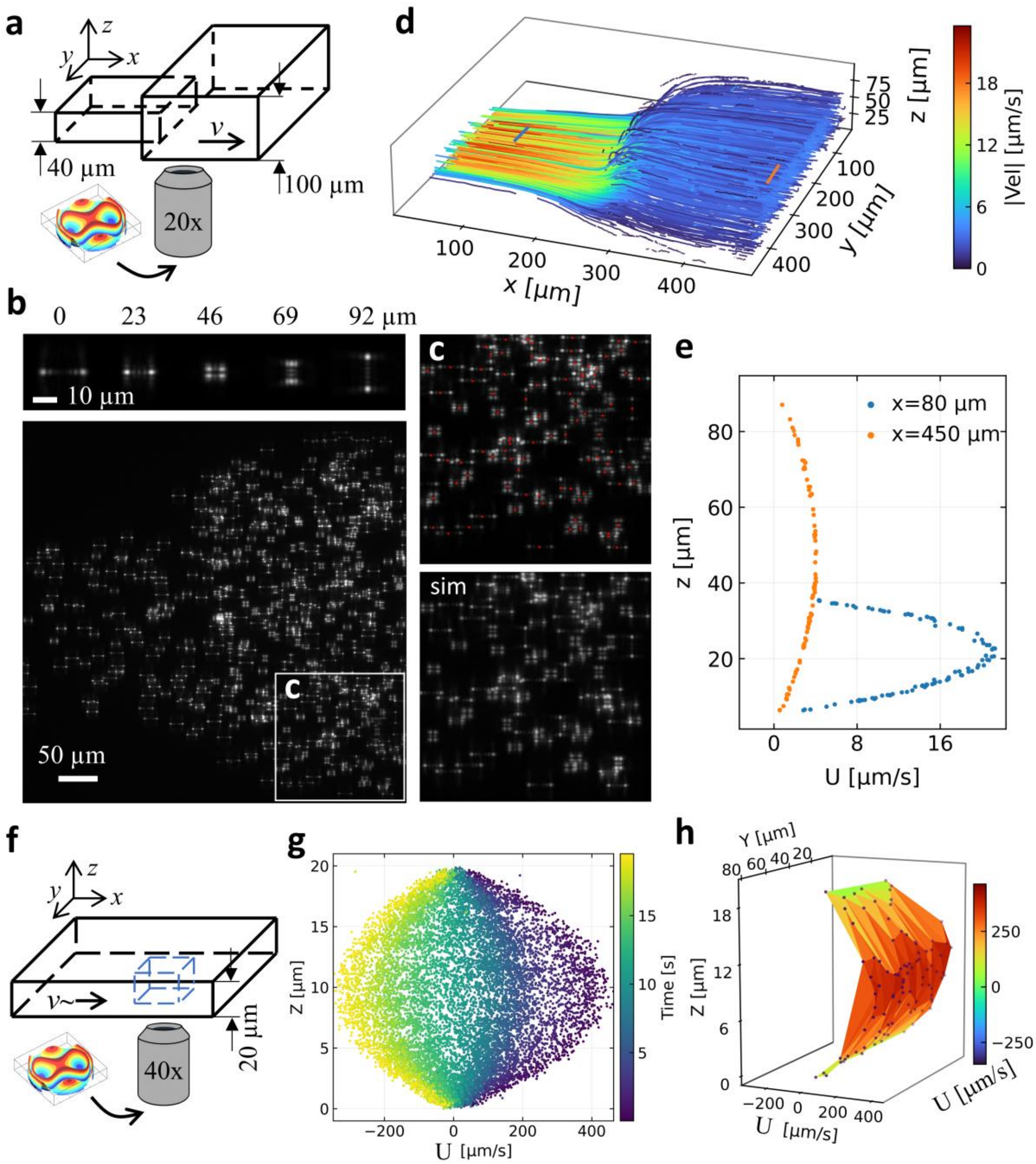

Fig. 2: Tracking mode of PET-PIV in microscopy. a-e, Steady flow measurement in a microchip with a cross-sectional area change. a, Imaging geometry. b, A representative frame with the calibrated PSF model on the top. c, Zoom-in view of region c in b showing x-y localizations with red marks in the raw image (top) and axial positions by reconstructed PSFs (bottom). The label "sim" denotes that the image is simulated, based on the found particle positions propagating through our imaging model. d, Particle trajectories colored by velocity magnitude. e, The distribution of x velocity component (U) along the depth (z) axis at the blue and orange line in d. f-h, Unsteady flow measurement in a straight microchannel. f, Imaging geometry. g, Time-varying U along depth axis. h, 2D U profile in the y–z plane at t = 0.2 s.

## Tracking mode of PET-PIV at the macroscale

At the macroscale, the PET-PIV system operates with a single camera (C0, Fig. 1) equipped with a Tetrapod phase mask mounted at the lens aperture. For validation purposes only, it is integrated with a TPIV system (Fig. 3a; see experimental system in Fig. S2c and Methods for physical dimensions and flow generation approaches) whose results serve as ground-truth references for the PET-PIV measurements.

First, sparse particle flow with minimal PSF overlap is imaged to calibrate camera C0 (see Methods section), and the resulting calibrated PSF model is employed to generate synthetic training images for LocNet. We evaluate network performance using centimeter-scale swirl flows at three seeding densities: 0.05, 0.1, and 0.2 PPCM (Supplementary Video S5, S6, and S7). At each density, image sequences from all five cameras are synchronously acquired. The C0 image sequence is processed by the trained LocNet to obtain the PET localization list, while the image sequences from cameras C1–C4 are processed using the STB algorithm to produce a reference localization list. PET localizations are then aligned to the STB coordinate system using a 3D affine transformation with an additional nonlinear correction term (see Methods section).

To quantify localization agreement, correspondences between the two localization sets are established by solving a linear sum assignment problem. Matching performance is quantified over 100 frames at each density using Dice similarity coefficient (DSC) and mean absolute error (MAE) (Fig. S4a). As seeding density increases, PSF overlap becomes more pronounced, making localization increasingly challenging. At 0.2 PPCM, a representative frame is shown in Fig. 3b, with an inset zoomed-in view and the calibrated PSF model displayed below. PET localizations in the zoomed region are presented in Fig. 3c: the measured image with x-y localizations (red marks) is shown on the left, and the simulated image illustrating z localization is shown on the right. A direct comparison between PET and STB localizations is provided in Fig. 3d (see 2D projections in Fig. S4b).

We use the intermediate-density case (0.1 PPCM) to further validate the linking and interpolation results. Ten-frame particle tracks colored by velocity magnitude are shown in Fig. 3e (see the corresponding STB result in Fig. S4c). The CCM interpolation algorithm is then applied to map the particle velocities onto an Eulerian grid, yielding volumetric velocity fields. An iso-surface comparison between PET and STB is presented in Fig. 3f. The reconstructed velocity components (U, V, W), together with their differences relative to STB, are shown with orthogonal slices in Fig. 3g. The difference is quantified with correlation coefficient (CC) which is ~0.98 on average (Fig. S4c) in this 100-frame dataset (full result in Supplementary Video S8).

Finally, we apply the tracking pipeline to a centimeter-scale jet-flow data (Supplementary Video S9) at the intermediate density with a Re of ~3000. The reconstructed velocity magnitudes from STB and PET are shown in Fig. 3h and Fig. 3i, respectively (see component-wise comparison in Fig. S5a and Supplementary Video S10), together with the absolute difference in Fig. 3j. A representative velocity-vector slices from the PET result is shown in Fig. S5b. The difference is CC of ~0.97 (Fig. S5d) on average in this 300-frame data.

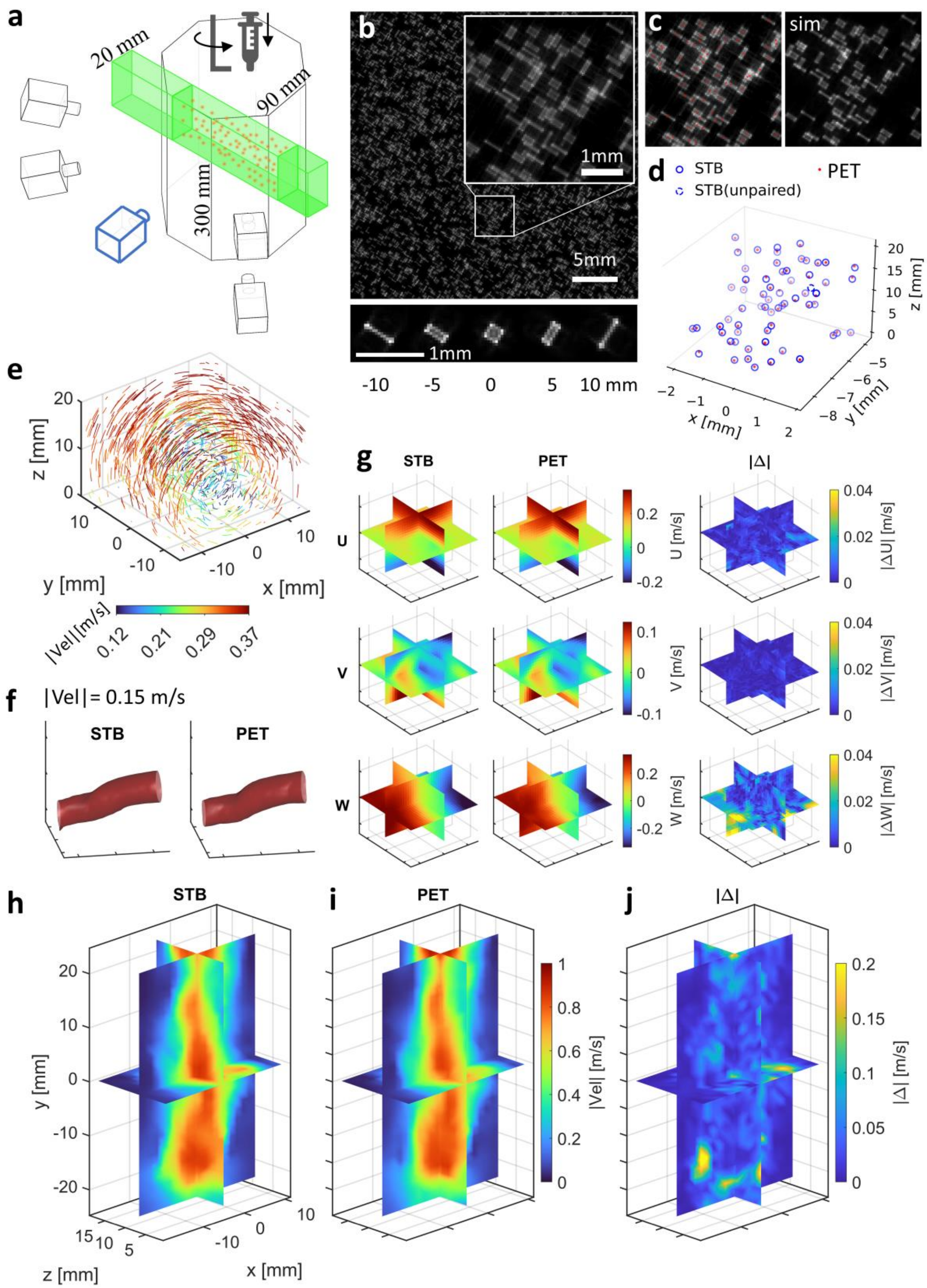

Fig. 3: Tracking mode of PET-PIV at the macroscale. a, Diagram of the experimental system where an Allen key attached to a screw driver and a syringe are used to generate swirl flows and jet flows

respectively. b, Representative high-density frame with inset zoomed-in view and the calibrated PSF model. c, Zoomed-in region showing x-y localizations (red marks, left) and simulated image illustrating z localization (right). d, 3D comparison of PET and STB localizations in the region c. e, Ten-frame PET particle tracks for the intermediate-density case. f, Comparison of velocity-magnitude iso-surfaces reconstructed from PET and STB. g, Velocity components (U, V, W) and their differences between PET and STB. h–j, Jet-flow validation: PET joint velocity magnitude (h), STB joint velocity magnitude (i), and their difference map (j).

## Field mode of PET-PIV

We demonstrate the field mode of PET-PIV in the same macro imaging system (Fig. 4a), albeit at higher seeding density. This mode directly maps a short sequence of consecutive 2D images to a 3D–3C velocity field using a neural network (VelNet; Fig. 4b), completely avoiding the particle localization and linking steps, because they are impractical at higher particle densities. Training is supervised using experimentally acquired datasets paired with aligned TPIV ground truth. We evaluated the trained VelNet on a 300-frame jet-flow video (Supplementary Video S11) with a seeding density of ~0.4 PPCM with a Re of ∼2000 (representative frame in Fig. 4c). At this density, TPIV is, to the best of our knowledge, the only feasible PIV approach, requiring a multi-camera configuration. PET-PIV agrees well with TPIV, yielding a mean CC of ∼0.98 for the velocity magnitude (histogram in Fig. 4e, component-wise CC analysis in Fig. S6b, and full result in Supplementary Video S12). A representative snapshot comparison of the three velocity components is shown in Fig. 4d. The per-frame processing times for TPIV and PET-PIV are around 30s and 1s, respectively.

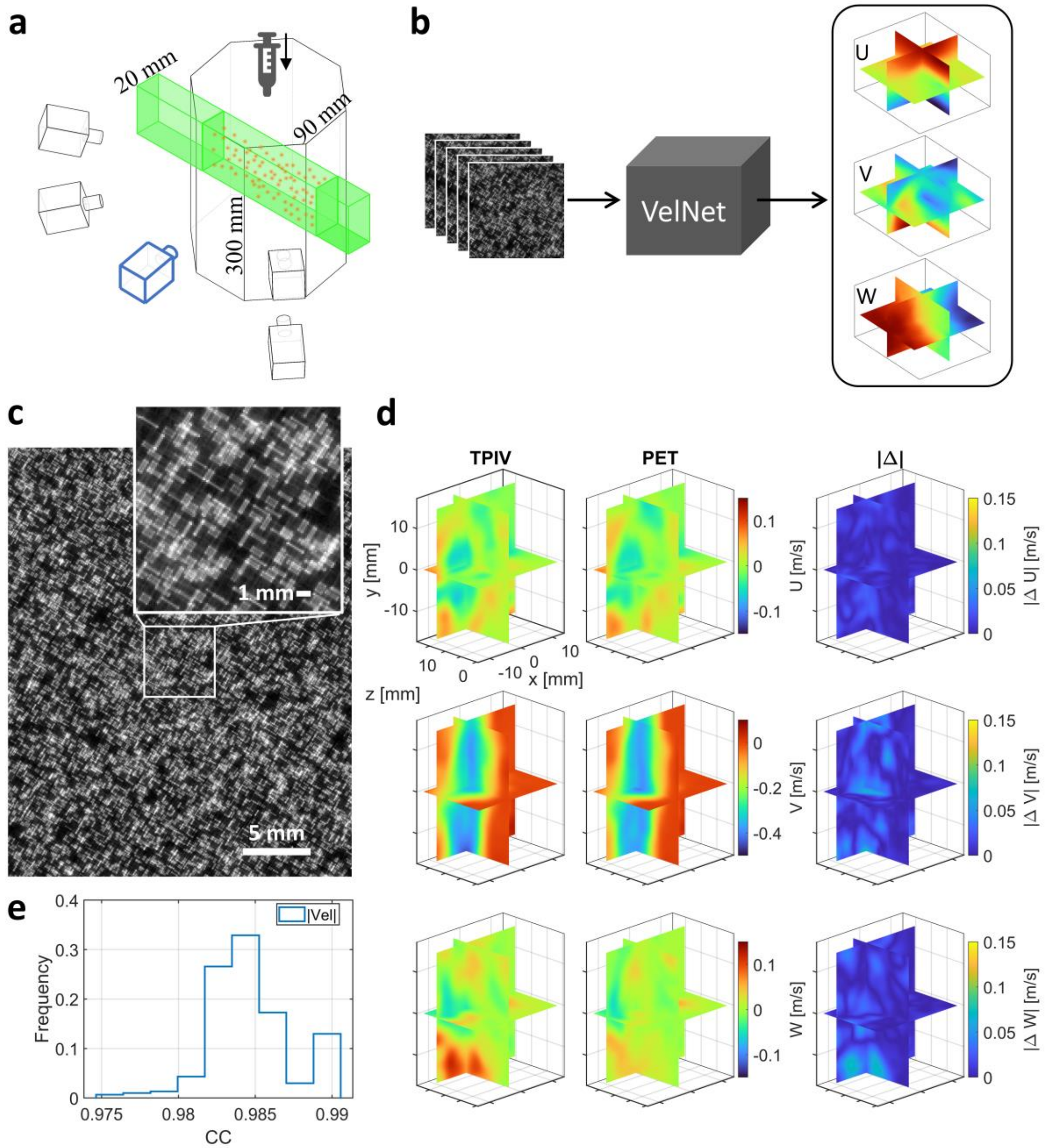

Fig. 4: Field mode of PET-PIV at the macroscale. a, Diagram of the imaging system where a syringe is used to generate jet flows. b, Schematic of the field mode, which takes a short sequence of consecutive 2D frames as input and directly predicts the 3D-3C velocity field using a neural network VelNet. c, A representative raw frame from a 300-frame recording at high particle density (~0.4 PPCM), with a zoomed-in inset. d, Orthogonal slices of the velocity components U, V, and W (rows 1–3) from TPIV (first column) and PET (second column), together with the corresponding difference maps (third column). Slices are taken at x=−3 mm, y=0 mm, and z=14 mm, approximately at the center of the jet flow. **e,** Histogram of the CC of velocity magnitude between VelNet predictions and TPIV ground truth; the mean CC is ~0.98.

## Discussion

We have demonstrated PET-PIV, a compact framework for volumetric velocimetry, capable of reconstructing the 3D distribution and motion of particles using a single camera, a simple phase mask, and learning-based reconstruction. The framework encodes depth-resolved particle information optically and recovers it computationally across imaging scales and particle densities. Across both microscopic and macroscopic imaging scales, PET-PIV achieves validated velocimetry performance, demonstrating strong agreement with the ground-truth reference at moderate and high densities across multiple realistic experimental benchmarks. The approach is readily deployable and scalable, requiring only a standard imaging system with a simple, easily implementable low-cost addition[28] (a single phase mask) together with our in-situ system calibration.

At the hardware level, PET-PIV exhibits almost trivial optical modification: in microscopic imaging, the phase mask is inserted directly beneath the objective[29], whereas in macro-scale imaging it is mounted at the lens iris[30]. This modification reshapes the particle image (PSF) such that depth information is encoded in the PSF footprint, gaining depth discrimination at the expense of an enlarged PSF.

The two operating regimes of PET-PIV address complementary limits of particulate imaging. The tracking mode provides particle-resolved information, including 3D positions, trajectories, and velocities. This regime is particularly important when the distribution and motion of individual particles, droplets, tracers, or particulate structures are themselves the quantities of interest. At ultra-high particle densities, where individual localization and linking become unreliable because of particle-image overlap, the field mode bypasses explicit tracking and maps image sequences directly to volumetric velocity fields. In the present work, this dense-regime reconstruction was validated against tomographic PIV, but the same conceptual framework could be extended to other volumetric quantities, including particle concentration, dispersion, residence time, or transport fields, provided that suitable calibration, simulation, or experimental ground truth is available.

A primary limitation of our current field-mode implementation is its reliance on supervised training with tomographic PIV ground truth. This requirement does not undermine the general framework, but it defines the present route by which high-density reconstructions are obtained. Future implementations may reduce this dependence through synthetic training data generated from coupled flow simulations and realistic particle-image models, through transfer learning between related flow classes, or through shared pretrained models for common measurement configurations. Similarly, the Tetrapod phase mask used here was chosen as a robust and practical depth-encoding element, but future systems could co-optimize the phase mask and reconstruction network for specific particles, imaging scales, density ranges, or target observables.

By combining wavefront-shaped monocular imaging with learning-based reconstruction, PET-PIV shifts volumetric particulate measurement from a specialized multi-camera infrastructure toward a compact and adaptable imaging platform. This could make 3D transport measurements accessible in confined geometries, field-deployable systems, biological and microfluidic environments, atmospheric and environmental flows, and industrial settings where optical access is limited. More broadly, the approach establishes a practical path for highly accessible imaging of how material is distributed and transported in three dimensions.

# Methods

## Flow systems

Microscopic PIV experiments were conducted in a system (Fig. S1f) consisting of a microscope (Nikon Eclipse Ti2), a 640-nm laser, a scientific CMOS camera (Photometrics Prime 95B, 10-μm pixel size), and a flow-control unit (Fluigent). A tetrapod phase mask (7.5-mm diameter), fabricated by photolithography, was mounted in a 3D-printed holder that fits the microscope's accessory slot (Fig. S1a). This compact insertion may increase field-dependent aberrations in the imaging system, which can be addressed computationally[31,32]. The same mask was used for both steady- and unsteady-flow experiments. The PSF model (Fig. S1c, d, e) was calibrated from a z-stack of fluorescent particles adhered to the bottom surface of the flow channel, although in situ calibration is also possible[33].

Fluorescent particles (excitation/emission: 625/645 nm; diameter: 1 μm) suspended in distilled water were used as tracers. For steady-flow experiments, we used a custom-designed passive microfluidic chip (Fig. S1b). For unsteady-flow experiments, we used a straight-channel microfluidic chip (ChipShop, product number: 10000216; channel height: 20 μm). The objectives used for the steady and unsteady experiments were 20× (NA 0.75) and 40× (NA 0.75), respectively.

Maco PIV experiments were performed in an octogen tank (side length: 90 mm; height: 300 mm), illuminated by a 527-nm dual-cavity laser (Photonics Industries). The setup (Fig. S2c) comprised four TPIV cameras (Photron FASTCAM Mini UX 100) equipped with Zeiss macro lenses (Milvus 100 mm f/2) and loss-pass filters (LP540), as well as a PET camera (Phantom v2640) equipped with a Tokina macro lens (ATX 100mm f2.8) and a long-pass filter (LP550). A custom-designed tetrapod phase mask (18-mm diameter), fabricated by photolithography, was placed at the aperture of the Tokina lens (Fig. S2b)[34].

The working fluid was distilled water seeded with fluorescent tracer particles (microParticles GmbH, PS-FluoRed-W62; diameter 25.56 μm; excitation/emission: 530/607 nm). Swirling flows were generated using an electric screwdriver, and jet flows were produced using a syringe (nozzle diameter: 3 mm) mounted at the top of the tank. Image sequences were acquired at 1 kHz for swirl experiments and 2 kHz for jet experiments.

## Imaging model

In microscopy, the imaging model is typically defined by the objective and is often formulated using Fraunhofer diffraction (optical Fourier transform). Deviations from the ideal imaging geometry, such as defocus, are incorporated through additional phase terms[35]. To bridge the gap between theoretical and experimental models, phase retrieval methods[36,37] based on z-stack measurements are widely used. Here, we analyze and solve a practical issue that can limit calibration accuracy: under-sampling.

When Fraunhofer diffraction is computed using the fast Fourier transform (FFT), the sampling relation is $N\Delta\xi = \frac{\lambda f}{\Delta x}$, where N is the simulation grid size, $\Delta\xi$ and $\Delta x$ are sampling intervals at the Fourier plane and image plane (camera pixel size divided by magnification), respectively, $\lambda$ is the wavelength, and $f$ is the focal length. The left side, $N\Delta\xi$, is the physical extent of the modeled Fourier plane. For large camera pixels and/or low magnification, this modeled extent can be smaller than the objective pupil, leading to the under-sampling issue and degraded phase-retrieval performance. A common remedy is to virtually decrease camera pixel size in the forward model, but this complicates implementation. Instead, we mitigate under-sampling by rescaling the pupil aperture in the model by a factor of $\frac{\lambda}{\Delta x NA} - 1$, which yielded good performance in our experiments.

For macro-scale PIV, we model the PSF of a particle located at (x, y, z) in object space, with the coordinate origin at the in-focus plane, as

$$PSF = \mathcal{N}\left(\mathcal{AF}_{s,r,x,y}\left(\left|\mathcal{F}\left(Ae^{i(M+czD)}\right)\right|^2\right)\right), \quad (1)$$

where M denotes phase mask, D is a parabolic defocus phase term, c is an approximated linear defocus coefficient, and A is the aperture. $\mathcal{F}$ denotes the Fourier transform operator, $\mathcal{AF}$ is an affine transform parameterized by scaling s, rotation r, translation x and y. $\mathcal{N}$ represents the noise model. This formulation[34] avoids the FFT sampling constraint, allowing the computational grid in the forward model to remain small. We implement the model in a fully differentiable manner, enabling retrieval of the unknown parameters via backpropagation.

## In-situ PSF calibration

Using sparse data without substantial PSF overlap, we performed in situ PSF calibration for camera C0 in the macro-scale PIV system. We first extracted isolated PSFs by applying Gaussian filtering followed by blob detection, yielding a pool of calibration PSFs (Fig. S3a). Based on the imaging model in Eq. (1), we then adopted an alternative and iterative optimization scheme to estimate two parameter sets: (i) the aperture phase M and (ii) per-particle properties, including 3D location and photon count.

Specifically, we initialized the forward model with the designed phase map of the inserted mask and with proper scaling and rotation factors. We then estimated the particle parameters by minimizing the Poisson negative log-likelihood using backpropagation and enforced an axial-range constraint determined by the laser sheet thickness (20 mm in our experiments). Next, with particle parameters fixed, we optimized the phase $M$ using mean squared error loss. We alternated between these two steps until convergence, which was typically achieved within a few iterations (a converged PSF model in Fig. S3c-e).

## Localization and linking

LocNet is a fully convolutional neural network adapted from localization microscopy[22,23,38]. It is trained on synthetic data generated using the calibrated PSF model. Each training sample comprises (i) an image containing particles randomly sampled in 3D space and (ii) a volumetric target in which each particle is represented by a 3D Gaussian blob, whose center of mass encodes the particle location. We used 10,000 image–label pairs for training, which typically required several hours.

During inference, LocNet maps an input image to a 3D volume. Particle positions are obtained by detecting the Gaussian blobs in the predicted volume and extracting their centers, yielding a localization list. For validation, the recovered localizations are passed through the forward imaging model to render a simulated image, which is compared to the measurement (Fig. 2c and Fig. 3c).

To generate tracks from the localization list, we used a four-frame linking algorithm adapted from[24]. The method constructs trajectories at the current time point by leveraging existing tracks from the previous frame together with localizations in the subsequent two frames. Because PET localizations are typically more precise in x-y than in z, the conventional 3D four-frame linking can be disrupted by the large axial uncertainty, an effect that becomes more pronounced at higher particle densities and leads to fragmented (short) tracks. To mitigate this issue, we performed the four-frame linking in the x-y plane only and enforced an additional bound on axial displacement to maintain consistency in z. For track visualizations, we used inTRACKtive[39].

## TPIV validation: Lagrangian ground truth via STB + CCM and Eulerian reference via 3D correlation

STB: Time-resolved particle tracking was performed using the Shake-The-Box (STB) algorithm[26] implemented in LaVision DaVis 11.2 (LaVision GmbH). STB reconstructs and tracks individual tracer particles in three dimensions by iteratively predicting particle motion and refining positions through multi-view reprojection. This predictor–corrector framework enables particle-resolved volumetric tracking at high seeding densities and serves as a reference for assessing 3D particle localization accuracy obtained from the single-camera PSF-based approach.

CCM: The resulting scattered particle velocities were converted to continuous velocity fields using the constrained cost minimization (CCM) framework[25]. In CCM, the Eulerian velocity field is reconstructed through global optimization that enforces consistency with measured particle velocities while imposing spatial regularity. This approach was adopted instead of local binning or kernel-based interpolation, which can become unreliable in regions with sparse particle sampling. The CCM formulation provides robust field reconstruction even when particle tracks are unevenly distributed, enabling direct comparison between particle-resolved and grid-based representations.

Independent volumetric velocity fields were reconstructed using tomographic particle image velocimetry[40], also implemented in LaVision DaVis 11.2 (LaVision GmbH). Particle intensity distributions were reconstructed from synchronized multi-view recordings using iterative algebraic reconstruction, followed by volumetric cross-correlation with multi-pass window refinement.

Tomo-PIV provides spatially uniform, grid-based velocity fields and serves two purposes in this study: (i) as supervised training data for the deep learning framework and (ii) as an independent Eulerian benchmark for evaluating volumetric flow fields reconstructed from the PSF-based single-camera system.

## Alignment between TPIV and PET-PIV

The alignment establishes a mapping from PET coordinates to TPIV coordinate system. Using sparse datasets, we first generated two localization lists, one from LocNet (PET) and one from STB (TPIV). We then progressively established correspondences by solving an assignment problem between the two sets of points and estimating a 3D affine transformation. To further improve accuracy, we fit a nonlinear interpolator to the residuals after the affine alignment, yielding a refined mapping from PET to the TPIV coordinates (Fig. S3f-g).

### VelNet

The VelNet architecture comprises four modules (Fig. S6a): depth decoding, cost volume computation, velocity prediction head, and a fuse gate. The depth-decoding module is adapted from LocNet and processes each input frame independently, mapping it to a 3D volume that represents particle occupancy in object space. Next, cost volumes[41] are computed at two temporal step sizes to capture 3D temporal-correlation features. The cost volumes from each step size are then passed to separate prediction heads: cost volumes at the first step size are used by Head 1 to predict a 3D–3C velocity field, while cost volumes at the second step size are used by Head 2 to produce an independent velocity estimate. Finally, the two velocity fields are fused using voxel-wise weights predicted by a gating network.

As an example, consider a three-frame input ($I_1$, $I_2$, $I_3$) and step sizes $\Delta t=1$ and $\Delta t=2$. The depth-decoding module converts each frame into a volume ($V_1$, $V_2$, $V_3$). Cost volumes at $\Delta t=1$ are computed between ($V_1$, $V_2$) and (V2, V3) and then are fed to Head 1 to produce a velocity field 1. In parallel, the cost volume at $\Delta t=2$ is computed between ($V_1$, $V_3$) and is fed to Head 2 to produce field 2. The final prediction is obtained by fusing the two predicted fields with voxel-wise gate weights. Notably, the cost-volume module is purely computational and contains no trainable parameters.

Training data were obtained from five jet-flow datasets: two acquired at a seeding density of 0.3 PPCM, one at 0.4 PPCM, and two at 0.5 PPCM. The 0.4-PPCM dataset was divided into two halves, with the second half used for training. For each dataset, the TPIV velocity fields was spatially aligned with the corresponding PET images. A velocity field was paired with a temporally centered image sequence spanning 5 frames. These aligned image–velocity pairs were then cropped into patches to form the final training samples.

Training proceeded in four stages. First, we initialized the depth-decoding module using a pretrained LocNet. Second, we froze the depth-decoding weights and trained only the two velocity-prediction heads and the fusion gate. Third, we froze these modules and fine-tuned the depth-decoding module. Finally, we jointly fine-tuned the entire VelNet end-to-end. An L1 loss was minimized at all stages. The first half of the 0.4-PPCM dataset was used to test the trained network.

## Acknowledgments

We would like to express our sincere gratitude to Dr. Ron Bessler for his assistance in managing the tomographic PIV system. We also thank ChatGPT for its support in developing the code for video demonstrations and for polishing the manuscript writing.

## Funding

Israel Science Foundation (grant No. 1081/24)

**Competing interests:** Authors declare that they have no competing interests.

**Data, code, and materials availability:** All data, code, and materials will be available in a GitHub repository.

## Bibliography

(1) Stone, H. A.; Stroock, A. D.; Ajdari, A. Engineering Flows in Small Devices: Microfluidics Toward a Lab-on-a-Chip. *Annual Review of Fluid Mechanics* **2004**, *36* (Volume 36, 2004), 381–411. https://doi.org/10.1146/annurev.fluid.36.050802.122124.
(2) Mo, H.; Hosni, M. H.; Jones, B. W. Application of Particle Image Velocimetry for the Measurement of the Airflow Characteristics in an Aircraft Cabin. *ASHRAE Transactions* **2003**, *109*, 101–110.
(3) Ragni, D.; van Oudheusden, B. W.; Scarano, F. 3D Pressure Imaging of an Aircraft Propeller Blade-Tip Flow by Phase-Locked Stereoscopic PIV. *Exp Fluids* **2012**, *52* (2), 463–477. https://doi.org/10.1007/s00348-011-1236-6.
(4) Raffel, M.; Bauknecht, A.; Ramasamy, M.; Yamauchi, G. K.; Heineck, J. T.; Jenkins, L. N. Contributions of Particle Image Velocimetry to Helicopter Aerodynamics. *AIAA Journal* **2017**, *55* (9), 2859–2874. https://doi.org/10.2514/1.J055571.
(5) Scarano, F. Tomographic PIV: Principles and Practice. *Meas. Sci. Technol.* **2012**, *24* (1), 012001. https://doi.org/10.1088/0957-0233/24/1/012001.
(6) Shi, S.; Ding, J.; Atkinson, C.; Soria, J.; New, T. H. A Detailed Comparison of Single-Camera Light-Field PIV and Tomographic PIV. *Exp Fluids* **2018**, *59* (3), 46. https://doi.org/10.1007/s00348-018-2500-9.
(7) Fahringer, T. W.; Lynch, K. P.; Thurow, B. S. Volumetric Particle Image Velocimetry with a Single Plenoptic Camera. *Meas. Sci. Technol.* **2015**, *26* (11), 115201. https://doi.org/10.1088/0957-0233/26/11/115201.
(8) Fort, C.; Solomon, H.; Capanna, R.; Wang, S.; Li, Z.; Bardet, P. M. Fourier Light-Field Microscope for Micro Tomo-PTV in a Cross-Junction Microfluidic Channel. *LXLASER* **2024**, *21*, 1–9. https://doi.org/10.55037/lxlaser.21st.195.
(9) Meng, H.; Pan, G.; Pu, Y.; Woodward, S. H. Holographic Particle Image Velocimetry: From Film to Digital Recording. *Meas. Sci. Technol.* **2004**, *15* (4), 673. https://doi.org/10.1088/0957-0233/15/4/009.
(10) Sheng, J.; Malkiel, E.; Katz, J. Digital Holographic Microscope for Measuring Three-Dimensional Particle Distributions and Motions. *Appl. Opt., AO* **2006**, *45* (16), 3893–3901. https://doi.org/10.1364/AO.45.003893.
(11) Xiong, J.; Idoughi, R.; Aguirre-Pablo, A. A.; Aljedaani, A. B.; Dun, X.; Fu, Q.; Thoroddsen, S. T.; Heidrich, W. Rainbow Particle Imaging Velocimetry for Dense 3D Fluid Velocity Imaging. *ACM Trans. Graph.* **2017**, *36* (4), 1–14. https://doi.org/10.1145/3072959.3073662.

(12) Xiong, J.; Aguirre-Pablo, A. A.; Idoughi, R.; Thoroddsen, S. T.; Heidrich, W. RainbowPIV with Improved Depth Resolution—Design and Comparative Study with TomoPIV. *Meas. Sci. Technol.* **2020**, *32* (2), 025401. https://doi.org/10.1088/1361-6501/abb0ff.
(13) Aguirre-Pablo, A. A.; Aljedaani, A. B.; Xiong, J.; Idoughi, R.; Heidrich, W.; Thoroddsen, S. T. Single-Camera 3D PTV Using Particle Intensities and Structured Light. *Exp Fluids* **2019**, *60* (2), 25. https://doi.org/10.1007/s00348-018-2660-7.
(14) Willert, C. E.; Gharib, M. Three-Dimensional Particle Imaging with a Single Camera. *Experiments in Fluids* **1992**, *12* (6), 353–358. https://doi.org/10.1007/BF00193880.
(15) Yoon, S. Y.; Kim, K. C. 3D Particle Position and 3D Velocity Field Measurement in a Microvolume via the Defocusing Concept. *Meas. Sci. Technol.* **2006**, *17* (11), 2897. https://doi.org/10.1088/0957-0233/17/11/006.
(16) Cierpka, C.; Segura, R.; Hain, R.; Kähler, C. J. A Simple Single Camera 3C3D Velocity Measurement Technique without Errors Due to Depth of Correlation and Spatial Averaging for Microfluidics. *Measurement Science and Technology* **2010**, *21* (4), 045401. https://doi.org/10.1088/0957-0233/21/4/045401.
(17) Cierpka, C.; Rossi, M.; Segura, R.; Kähler, C. J. On the Calibration of Astigmatism Particle Tracking Velocimetry for Microflows. *Measurement Science and Technology* **2011**, *22* (1), 015401. https://doi.org/10.1088/0957-0233/22/1/015401.
(18) Brockmann, P.; Hussong, J. On the Calibration of Astigmatism Particle Tracking Velocimetry for Suspensions of Different Volume Fractions. *Exp Fluids* **2021**, *62* (1), 23. https://doi.org/10.1007/s00348-020-03120-4.
(19) Shechtman, Y.; Sahl, S. J.; Backer, A. S.; Moerner, W. E. Optimal Point Spread Function Design for 3D Imaging. *Phys. Rev. Lett.* **2014**, *113* (13), 133902. https://doi.org/10.1103/PhysRevLett.113.133902.
(20) Shechtman, Y.; Weiss, L. E.; Backer, A. S.; Sahl, S. J.; Moerner, W. E. Precise Three-Dimensional Scan-Free Multiple-Particle Tracking over Large Axial Ranges with Tetrapod Point Spread Functions. *Nano Letters* **2015**, *15* (6), 4194–4199. https://doi.org/10.1021/acs.nanolett.5b01396.
(21) Cierpka, C.; Kähler, C. J. Particle Imaging Techniques for Volumetric Three-Component (3D3C) Velocity Measurements in Microfluidics. *J Vis* **2012**, *15* (1), 1–31. https://doi.org/10.1007/s12650-011-0107-9.
(22) Nehme, E.; Freedman, D.; Gordon, R.; Ferdman, B.; Weiss, L. E.; Alalouf, O.; Naor, T.; Orange, R.; Michaeli, T.; Shechtman, Y. DeepSTORM3D: Dense 3D Localization Microscopy and PSF Design by Deep Learning. *Nat Methods* **2020**, *17* (7), 734–740. https://doi.org/10.1038/s41592-020-0853-5.
(23) Saguy, A.; Xiao, D.; Narayanasamy, K. K.; Nakatani, Y.; Gustavsson, A.-K.; Heilemann, M.; Shechtman, Y. One-Click Image Reconstruction in Single-Molecule Localization Microscopy via Deep Learning. bioRxiv April 18, 2025. https://doi.org/10.1101/2025.04.13.648574.
(24) Shnapp, R. MyPTV: A Python Package for 3D Particle Tracking. *Journal of Open Source Software* **2022**, *7* (75), 4398. https://doi.org/10.21105/joss.04398.
(25) Agarwal, K.; Ram, O.; Wang, J.; Lu, Y.; Katz, J. Reconstructing Velocity and Pressure from Noisy Sparse Particle Tracks Using Constrained Cost Minimization. *Exp Fluids* **2021**, *62* (4), 75. https://doi.org/10.1007/s00348-021-03172-0.
(26) *Shake-The-Box: Lagrangian particle tracking at high particle image densities | Experiments in Fluids*. https://link.springer.com/article/10.1007/s00348-016-2157-1 (accessed 2024-10-28).
(27) Zimmermann, M.; Schmid, H.; Hunziker, P.; Delamarche, E. Capillary Pumps for Autonomous Capillary Systems. *Lab Chip* **2006**, *7* (1), 119–125. https://doi.org/10.1039/B609813D.

(28) Orange kedem, R.; Opatovski, N.; Xiao, D.; Ferdman, B.; Alalouf, O.; Kumar Pal, S.; Wang, Z.; von der Emde, H.; Weber, M.; Sahl, S. J.; Ponjavic, A.; Arie, A.; Hell, S. W.; Shechtman, Y. Near Index Matching Enables Solid Diffractive Optical Element Fabrication via Additive Manufacturing. *Light Sci Appl* **2023**, *12* (1), 222. https://doi.org/10.1038/s41377-023-01277-1.

(29) Baddeley, D.; Cannell, M. B.; Soeller, C. Three-Dimensional Sub-100 Nm Super-Resolution Imaging of Biological Samples Using a Phase Ramp in the Objective Pupil. *Nano Res.* **2011**, *4* (6), 589–598. https://doi.org/10.1007/s12274-011-0115-z.

(30) Levin, A.; Fergus, R.; Durand, F.; Freeman, W. T. Image and Depth from a Conventional Camera with a Coded Aperture. *ACM Trans. Graph.* **2007**, *26* (3), 70-es. https://doi.org/10.1145/1276377.1276464.

(31) Ferdman, B.; Saguy, A.; Xiao, D.; Shechtman, Y. Diffractive Optical System Design by Cascaded Propagation. *Opt. Express, OE* **2022**, *30* (15), 27509–27530. https://doi.org/10.1364/OE.465230.

(32) Xiao, D.; Kedem Orange, R.; Opatovski, N.; Parizat, A.; Nehme, E.; Alalouf, O.; Shechtman, Y. Large-FOV 3D Localization Microscopy by Spatially Variant Point Spread Function Generation. *Science Advances* **2024**, *10* (10), eadj3656. https://doi.org/10.1126/sciadv.adj3656.

(33) Liu, S.; Chen, J.; Hellgoth, J.; Müller, L.-R.; Ferdman, B.; Karras, C.; Xiao, D.; Lidke, K. A.; Heintzmann, R.; Shechtman, Y.; Li, Y.; Ries, J. Universal Inverse Modeling of Point Spread Functions for SMLM Localization and Microscope Characterization. *Nat Methods* **2024**, *21* (6), 1082–1093. https://doi.org/10.1038/s41592-024-02282-x.

(34) Xiao, D.; Kedem, R. O.; Shechtman, Y. Point Spread Function Modeling and Engineering in Black-Box Lens Systems. *Opt. Express, OE* **2025**, *33* (3), 4211–4224. https://doi.org/10.1364/OE.546767.

(35) Hanser, B. M.; Gustafsson, M. G. L.; Agard, D. A.; Sedat, J. W. Phase-Retrieved Pupil Functions in Wide-Field Fluorescence Microscopy. *Journal of Microscopy* **2004**, *216* (1), 32–48. https://doi.org/10.1111/j.0022-2720.2004.01393.x.

(36) Shechtman, Y.; Eldar, Y. C.; Cohen, O.; Chapman, H. N.; Miao, J.; Segev, M. Phase Retrieval with Application to Optical Imaging: A Contemporary Overview. *IEEE Signal Processing Magazine* **2015**, *32* (3), 87–109. https://doi.org/10.1109/MSP.2014.2352673.

(37) Xiao, D.; Ye, Q.; Tong, Z.; Xiang, B.; Wang, N. Exact Phase Recovery Applying Only Phase Modulations in an Isolated Region. *Optics and Lasers in Engineering* **2020**, *134*, 106278. https://doi.org/10.1016/j.optlaseng.2020.106278.

(38) Goldenberg, O.; Xiao, D.; Shechtman, Y. Deep Learning for Localization Microscopy in 2D and 3D. *Acc. Chem. Res.* **2026**. https://doi.org/10.1021/acs.accounts.6c00345.

(39) Huijben, T. A. P. M.; Anderson, A. G.; Sweet, A.; Hoops, E.; Larsen, C.; Awayan, K.; Bragantini, J.; Chiu, C.-L.; Royer, L. A. inTRACKtive — A Web-Based Tool for Interactive Cell Tracking Visualization. bioRxiv October 20, 2024, p 2024.10.18.618998. https://doi.org/10.1101/2024.10.18.618998.

(40) Elsinga, G. E.; Scarano, F.; Wieneke, B.; van Oudheusden, B. W. Tomographic Particle Image Velocimetry. *Exp Fluids* **2006**, *41* (6), 933–947. https://doi.org/10.1007/s00348-006-0212-z.

(41) Sun, D.; Yang, X.; Liu, M.-Y.; Kautz, J. PWC-Net: CNNs for Optical Flow Using Pyramid, Warping, and Cost Volume; IEEE Computer Society, 2018; pp 8934–8943. https://doi.org/10.1109/CVPR.2018.00931.

## Appendices

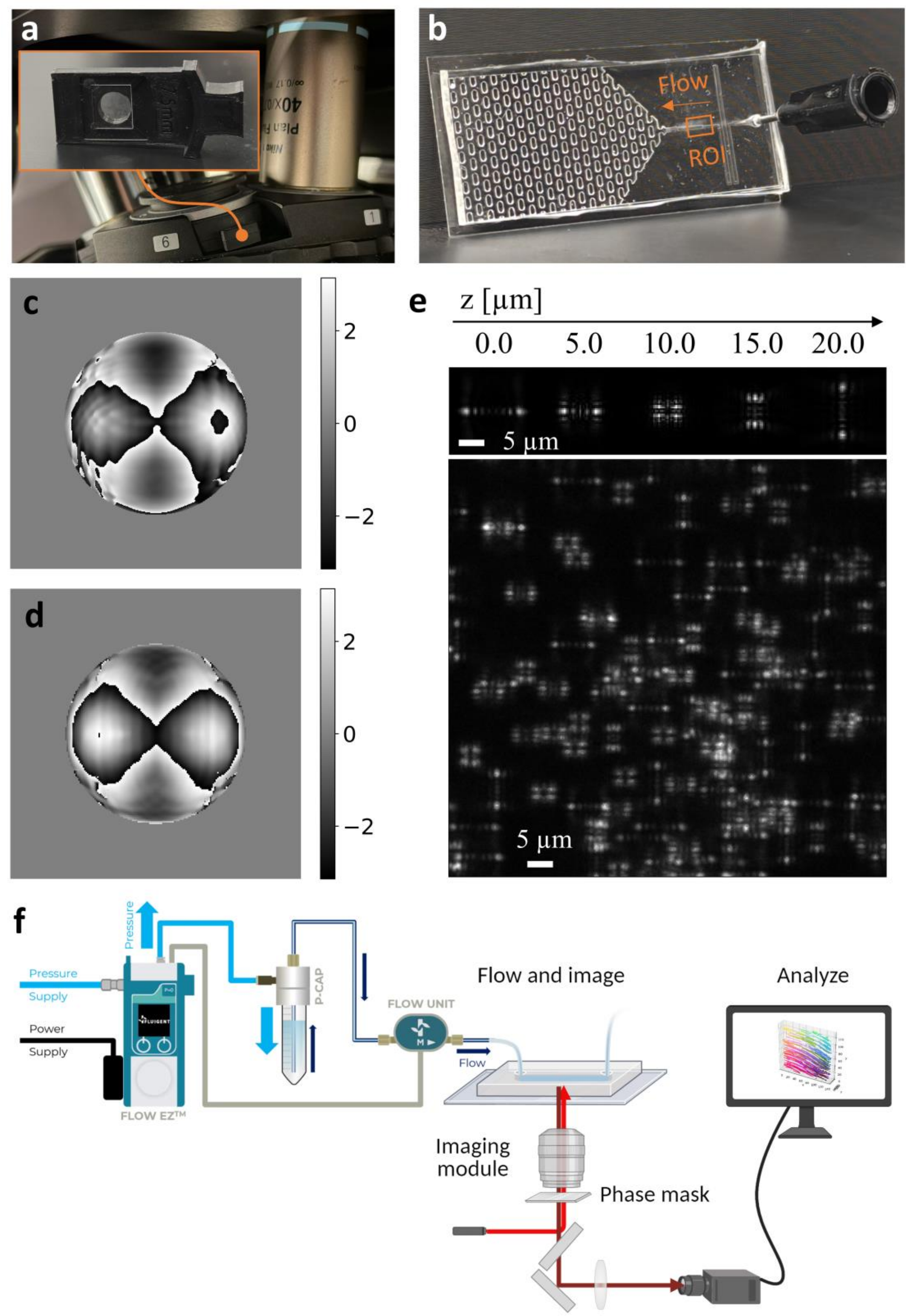

Fig. S1: PET PIV in microscopy. a, Phase-mask insertion beneath the microscope objective. b, Passive microfluidic chip used for steady-flow experiments in Fig. 2a. c, Retrieved pupil phase map

for the configuration in Fig. 2a. d, Retrieved pupil phase map for the configuration in Fig. 2f. e, Calibrated PSF model and a representative raw image from the experiment in Fig. 2f. f, Flow control system used in Fig. 2f.

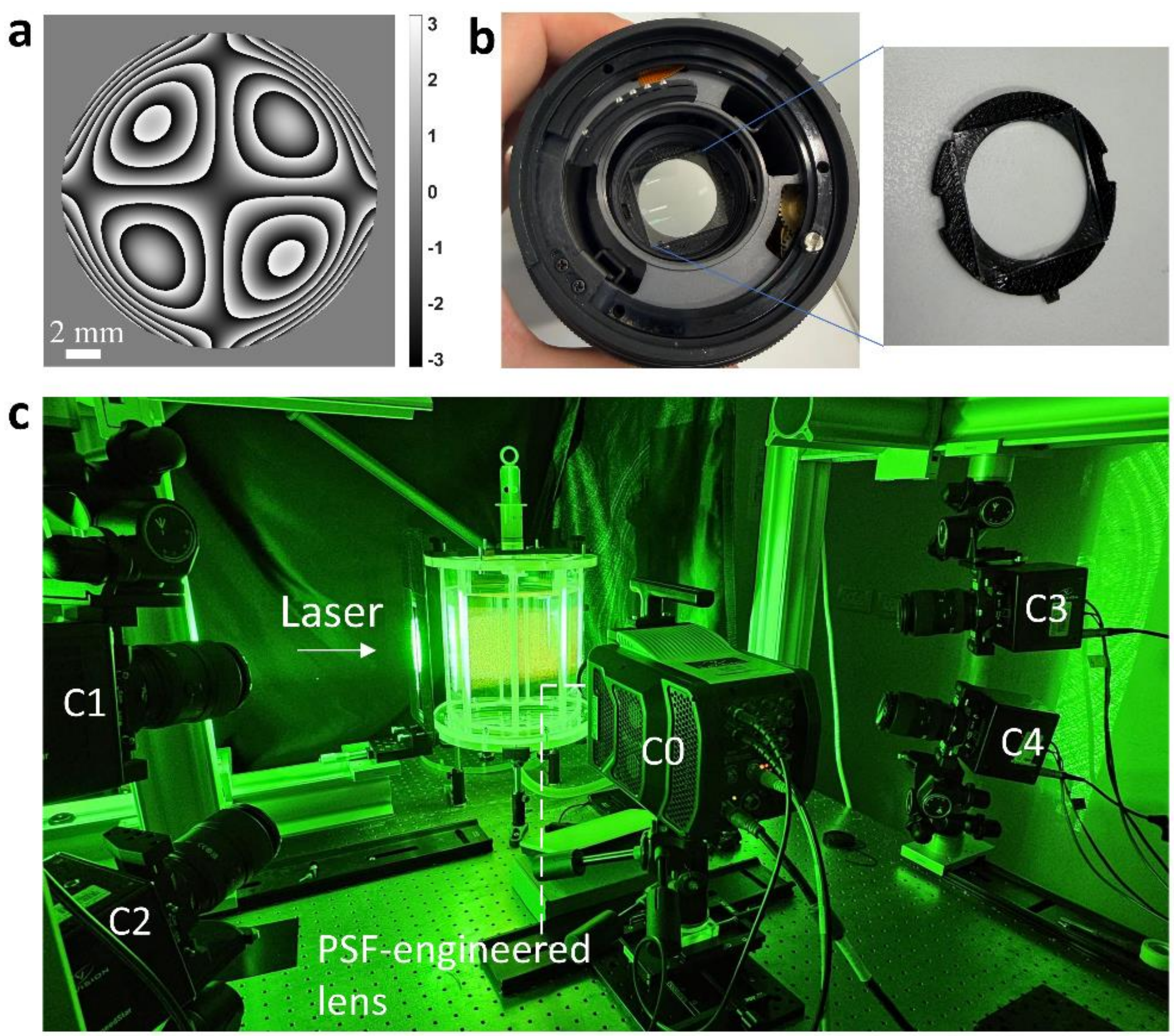

Fig. S2: TPIV and PET PIV in macro imaging. a, Designed phase map of the mask used in the PET camera (C0). b, Phase-mask insertion at the iris of a macro lens. c, Integration of PET-PIV into the TPIV system.

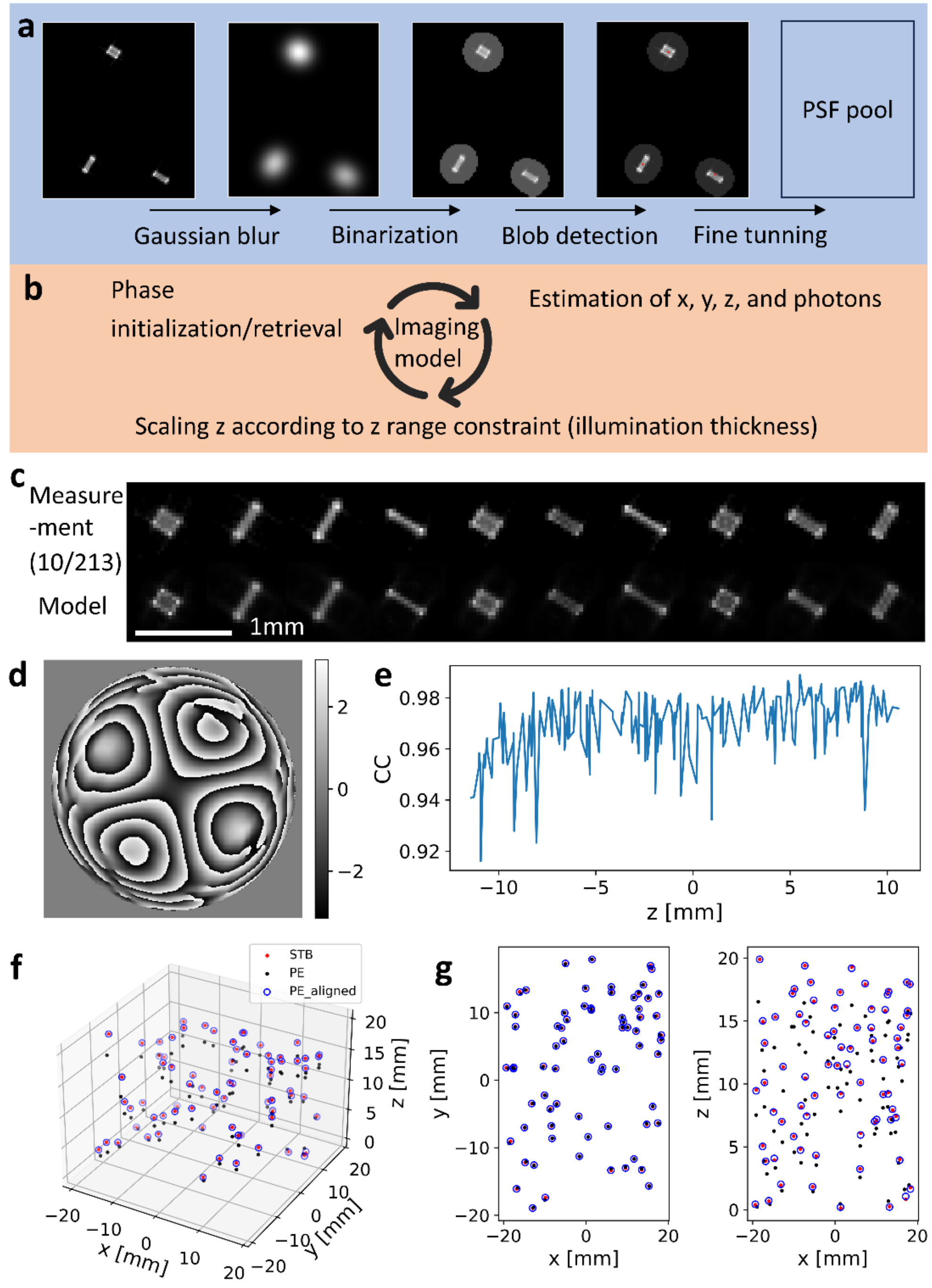

Fig. S3: In situ PSF calibration and alignment between TPIV and PET-PIV. a, Extraction of experimental PSFs from sparse flow data using Gaussian smoothing, binarization, blob detection, and refinement to form a calibration pool. b, Alternating optimization of the pupil phase and particle parameters based on the forward imaging model, with an imposed axial-range constraint. c, Representative examples (10 of 213) comparing measured PSFs with the corresponding model predictions. d, In situ–calibrated pupil phase map. e, Correlation coefficient (CC) between measured and modeled PSFs as a function of the estimated axial position. f, 3D alignment between STB and PET localizations. g, x-y and x-z projections of the alignment shown in f.

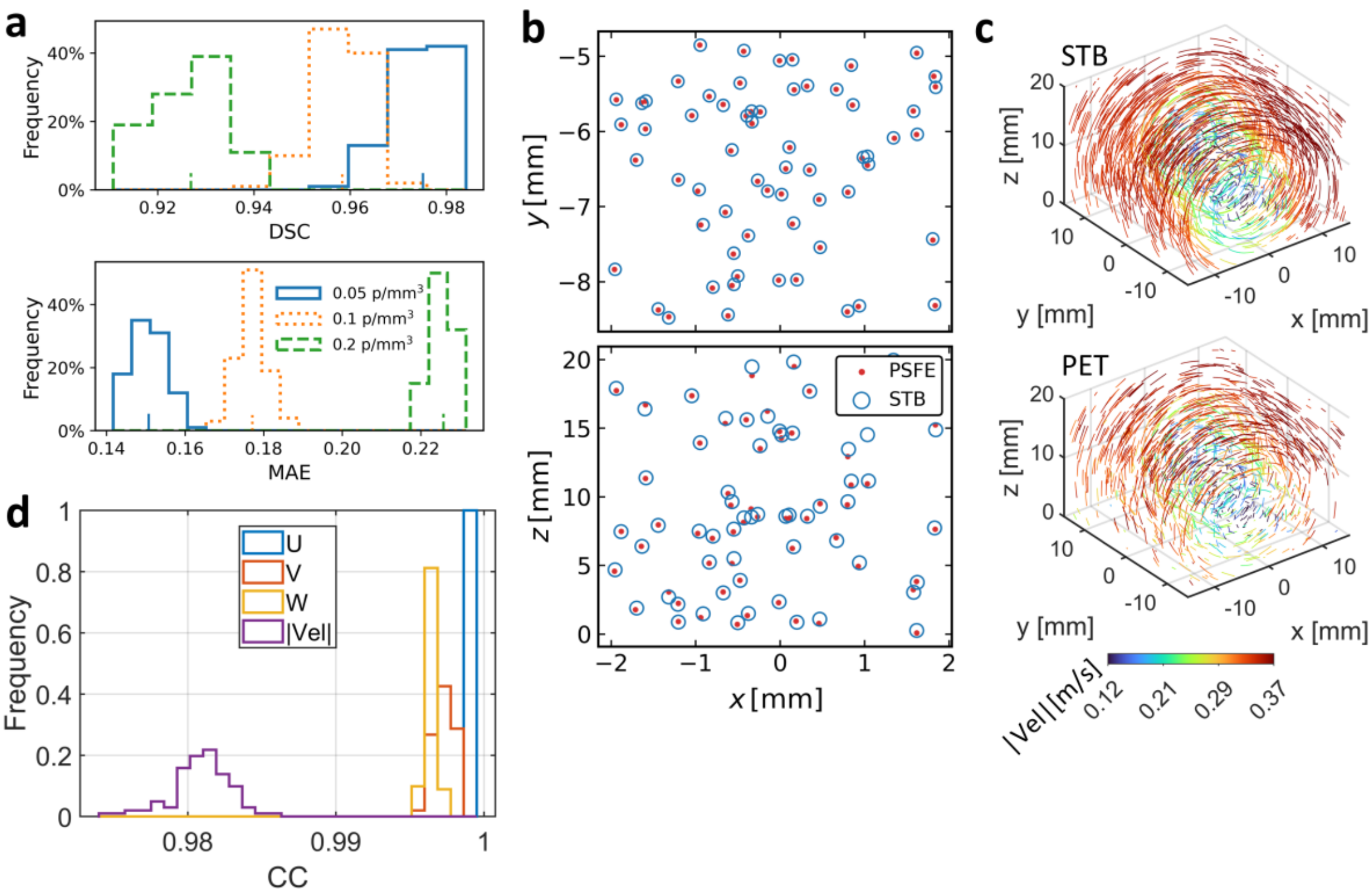

Fig. S4: Additional localization validation results and swirl results of PET-PIV tracking-mode in macro-scale imaging. a, Localization performance quantified by 100-frame histogram of DSC and MAE at three seeding densities. b, x-y and x-z projections for the localization-validation result in Fig. 2d. c, Trajectory comparison between STB and PET in the swirl flow in Fig. 2e. d, CC histograms of the swirl flow in Fig. 2.

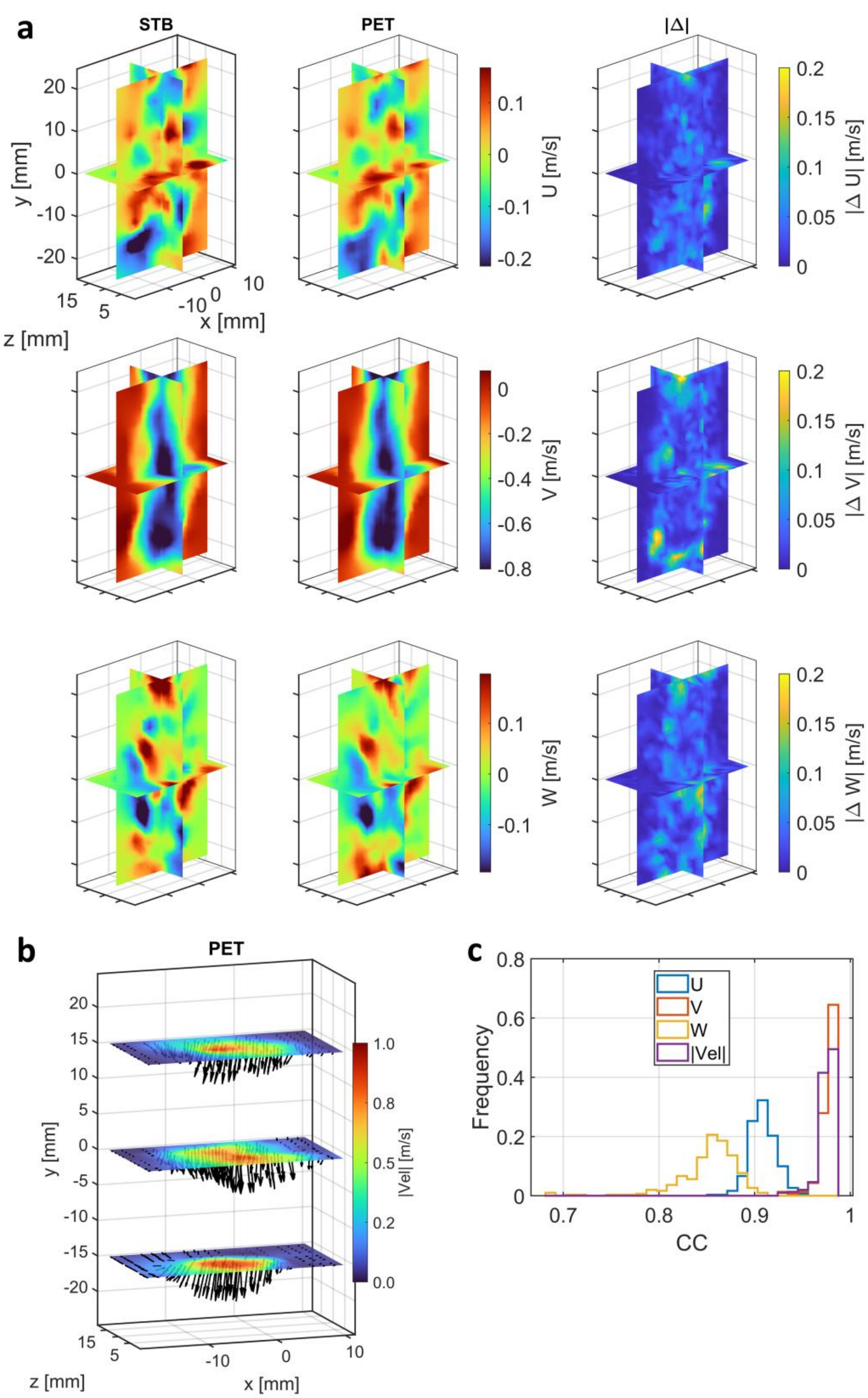

Fig. S5: Additional jet results of PET-PIV tracking-mode in macro-scale imaging. a, Component-wise velocity comparison for the jet flow shown in Fig. 2h–i. b, velocity vector plot at three x-z slices with color-coded velocity magnitude. d, CC histograms of the jet flow in Fig. 2.

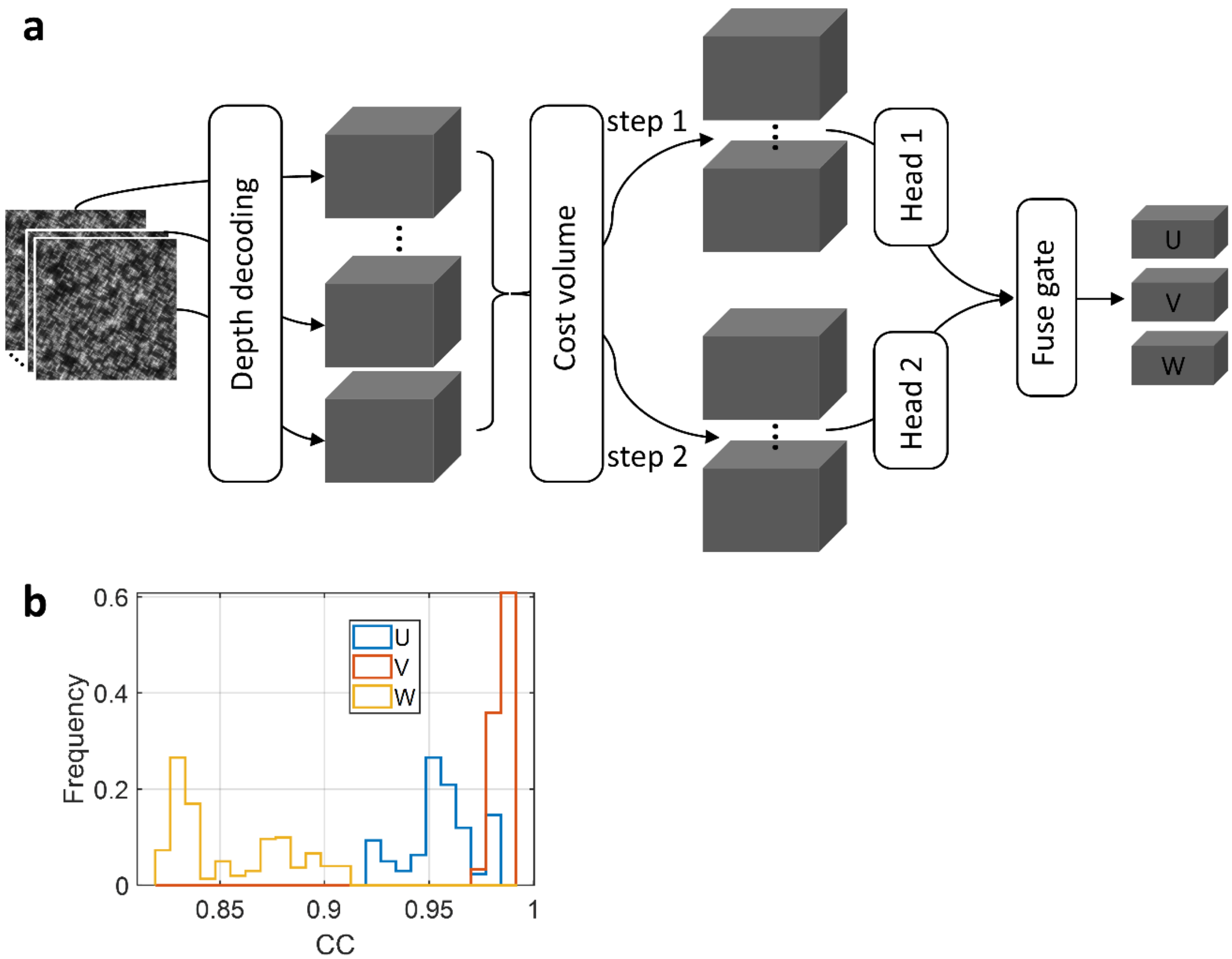

Fig. S6: VelNet architecture and component-wise correlation analysis. a, VelNet architecture comprising four modules: depth decoding, cost-volume computation, velocity prediction heads, and a fusion gate. b, CC histograms of U, W, and W corresponding to Fig. 4c.